\documentclass[preprint]{aastex}

\shorttitle{Evidence for Extinction and Accretion Variability in T Tau S}
\shortauthors{Beck, Prato \& Simon}

\begin{document}


\title{Evidence for Extinction and Accretion Variability in T Tau S}


\author{Tracy L. Beck\altaffilmark{1,2}, L. Prato\altaffilmark{3} and M. Simon\altaffilmark{1,2}}

\email{tracy@hilo.ess.sunysb.edu}

%\and

%\author{R. J. Hanisch\altaffilmark{5}}
%\affil{Space Telescope Science Institute, Baltimore, MD 21218}


\altaffiltext{1}{Department of Physics \& Astronomy, SUNY Stony Brook, Stony Brook, NY 11794-3800}
\altaffiltext{2}{Visiting Astronomer at the Infrared Telescope Facility, which is operated by the University of Hawaii under contract to the National Aeronautics and Space Administration.}
\altaffiltext{3}{Department of Physics \& Astronomy, University of California, Los Angeles, Los Angeles, CA 90095-1562}

%\altaffiltext{2}{Society of Fellows, Harvard University.}
%\altaffiltext{3}{present address: Center for Astrophysics,
%    60 Garden Street, Cambridge, MA 02138}
%\altaffiltext{4}{Visiting Programmer, Space Telescope Science Institute}
%\altaffiltext{5}{Patron, Alonso's Bar and Grill}


\begin{abstract}

We present angularly resolved spectra of T Tau North and South in the 3 $\mu$m water ice feature and K-band. Most of the water ice absorption lies along the line of sight toward T Tau South, confirming that it is viewed through stronger extinction.  A decrease in ice-band absorption toward T Tau S between December 1998 and January 2000, significant at the 2$\sigma$ level, was associated with an increase in its near infrared flux.   Br$\gamma$ emission is detected in T Tau North and South and H$_{2}$ (2.12 $\mu$m) emission only toward T Tau South, consistent with previous studies of infrared companions to T Tauri stars.  Our results suggest that the near IR variability of T Tau S is probably caused by both variations in accretion rate and variable extinction along the line of sight.

\end{abstract}


\keywords{stars: pre-main sequence, stars: individual (T Tau)}


\section{Introduction}

T Tau is the eponymous pre-main sequence star (Joy 1945), but observational investigations over the last 20 years have revealed that it is an exceptional member of the class.  Many of the young stars in the Taurus dark clouds are multiple systems, and in this respect T Tau is not atypical (Ghez et al. 1993; Leinert et al. 1993; Simon et al. 1995).  It consists of T Tau North (T Tau N), the star seen in visible light at V $\sim$10 (Herbig \& Bell 1988), and T Tau South (T Tau S), discovered in the infrared by Dyck, Simon \& Zuckerman (1982), at a projected separation $\sim$100 AU (0.$''$7).  T Tau S is itself a binary, with a projected separation of 7 AU (0.$''$05) (Koresko 2000).  Our observations do not resolve the 0.$''$05 T Tau S binary, so for the purpose of this paper we will refer to it as T Tau S. 

The T Tau system is remarkable in part because the physical nature of T Tau S is unknown.  It dominates the total system flux in the thermal IR, has a total luminosity two times that of T Tau N (Ghez et al. 1991, hereafter G91), but has never been detected in visible light to limiting V magnitude of 19.6 (Stapelfeldt et al. 1998).  T Tau S also shows pronounced variability in the infrared.  G91 discovered that it flared by $\sim$ 2 magnitudes in the near and thermal IR in 1990.  They concluded that T Tau S experienced enhanced accretion, similar to an FU Orionis outburst.  Kobayashi et al. (1994) and Simon et al. (1996) reported that the K band flux of T Tau S had decreased to the pre-flare value in the early 1990's. 

T Tau S is now regarded as a member of a small class of stars known as the infrared companions (IRCs)(Koresko et al. 1997, hereafter K97).  IRCs are companions to visible T Tauri stars, are very luminous in the IR, and show pronounced variability.   The near IR colors of IRCs imply a color temperature $<$ 1500 K, lower than possible for a stellar photosphere.  K97 suggest that the IRCs are young stars seen through strong extinction and are experiencing episodic accretion from their complex surrounding environment.

Spectral diagnostics in the near infrared provide probes to study extinction and accretion in IRCs.  Whittet et al. (1988) reported that the optical depth of the 3$\mu$m ice-band absorption feature, $\tau_{ice}$, correlates well with extinction in visible light, A$_{v}$. Sato et al. (1990) discovered the 3 $\mu$m ice feature in absorption toward the unresolved T Tau system.  Leinert et al. (1996)  monitored the ice-band optical depth toward the Haro 6-10 binary, an IRC system similar to T Tau, and found that both components varied in $\tau_{ice}$, possibly attributable to variations in extinction along the line of sight. 

Herbst et al. (1995; 1996; 1997) detected H$_{2}$ emission in the {\it v}=1-0 S(1) line (2.12 $\mu$m) in the Haro 6-10 and UY Aur IRCs.  The Herbst et al. (1996) observation of T Tau indicated peaks of unresolved emission at the positions of T Tau N and S.  The absence of emission in the {\it v}=2-1 S(1) (2.25 $\mu$m) transition in these sources suggests a line ratio consistent with a shock origin rather than fluorescence.  The observed H$_{2}$ emission could be caused by accretion shocks from the infall of material onto the star.   Prato \& Simon (1997) used Br$\gamma$ line emission as a surrogate for H$\alpha$ to distinguish between the classical and weak-lined T Tauri stars and Muzzerole, Hartmann \& Calvet (1998) showed that the Br$\gamma$ line luminosity in classical T Tauris correlates well with accretion luminosity determined from blue continuum excess.

To determine if the observed variability of T Tau S could be caused by variable extinction, we made angularly resolved observations of the near IR spectral energy distribution (SED) of T Tau N and S to test for variation in ice-band absorption on three occasions.  Additionally, we obtained angularly resolved K-band spectra of T Tau N and S to study Br$\gamma$ and H$_{2}$ accretion diagnostics, and if possible, to detect evidence of the photosphere of T Tau S. 

\section{Observations}

Ice-band observations were taken on UT 1998 December 11, 1999 November 17, and 2000 January 27 at NASA's Infrared Telescope Facility (IRTF) during scheduled observing time (December) and as part of the IRTF service observing program (November \& January).  We used NSFCam, the facility near IR camera, which is equipped with a 256 $\times$ 256 InSb array.  All of our observations were taken with the 0.$''$056/pixel plate scale.  To measure the spectral energy distribution (SED) in the ice-band, we used the K (2.2 $\mu$m) and L$'$ (3.8 $\mu$m)  broadband filters and the circularly variable filter (CVF), which has a spectral band pass of 2\%, set at 2.4, 2.9, 3.05 and 3.4 $\mu$m.  In December 1998 we also used a CVF setting at 3.6 $\mu$m and the H-band (1.65 $\mu$m) filter.  To freeze the seeing, we took 100 0.1 second exposures in the K and L$'$ filters, 200 0.1 sec exposures at all of the CVF filters except 2.9 $\mu$m.  The relatively weak flux of T Tau S in the H band and atmospheric absorption at 2.9 $\mu$m required 400 0.1 sec exposures.  

Figure 1 is a contour plot of T Tau N and S at 3.05 $\mu$m, demonstrating the resolution of the system in the ice-band.  16 Tau and $\eta$ Tau were the photometric standards for the broadband and CVF observations, respectively.  The zero-point fluxes for the CVF observations were determined using a polynomial fit to the broadband IR fluxes of Vega, as reported in Table 7.5 of {\it Allen's Astrophysical Quantities} (Tokunaga 1999).  We determined the Point Spread Function (PSF) at each of the observed wavelengths by making shift-and-add (S/A) images from a series of short exposures of the photometric standards.  The flux ratios of T Tau N and S at the observed wavelengths were derived using two different binary fitting procedures; the results were always consistent to within 1$\sigma$.  A discussion of the binary fitting routines will be included in a forthcoming paper (Beck et al., in preparation).

Our K-band (2.05 to 2.4 $\mu$m) spectral observations of the T Tau system were made at the Univerisity of Hawaii 2.2 meter telescope on UT 1996 November 2 with KSPEC, the facility near IR spectrometer (Hodapp et al. 1994; 1996).  This work closely followed S/A observations of T Tau in the K-band on UT 1996 October 26 at the IRTF which were subsequently used to flux-calibrate the KSPEC spectra.  The KSPEC detector is a 1024 $\times$ 1024 pixel HgCdTe array with a plate scale of 0.$''$167/pixel.  We used the tip-tilt adaptive optics system with the telescope in its f/31 configuration.  With a slit width of 0.$''$8, we obtained a resolution of $\sim$760 at 2.2 $\mu$m.  The slit was aligned at the position angle of the T Tau N-S system so that we acquired spectra of both stars simultaneously in a 35 minute exposure.

We observed the spectrum of HR 1377, a B8 single star, immediately following T Tau to provide PSFs in the cross-dispersion direction as well as a reference spectrum to correct that of T Tau for telluric absorption and instrumental effects.  The spectra of T Tau N and S were extracted by fitting, at each wavelength, model binaries constructed from the PSF of HR 1377, constrained by knowledge of the binary separation.  The total amplitude and flux ratio of the binary model which yielded the best fit in the least squares sense was used to derive the individual spectral amplitudes of T Tau N and S at each wavelength.  The spectra of T Tau N and S were divided by that of HR 1377 with the effects of its broad Br$\gamma$ absorption removed following the procedure described by Greene \& Meyer (1995). In order to restore the true form of the spectrum,  we multiplied by a black-body curve of temperature 11,400 K, that of a B8 star (Drilling \& Landolt 1999).  

The flux ratio of T Tau S to N of these spectra at 2.2 $\mu$m, 0.18$\pm$0.01, is in excellent agreement with the value of 0.16$\pm$0.03 measured by the S/A observations in the K-band six days earlier at the IRTF.  We applied the IRTF flux calibration to the spectra and show the final results in Figure 2.  Figure 3 is an expanded plot of the 2.10 to 2.20 $\mu$m region, which also shows a polynomial fit to the continuum.

%% In this section, we use  the \subsection command to set off
%% a subsection.  \footnote is used to insert a footnote to the text.


%% This section also includes several of the displayed math environments
%% mentioned in the Author Guide.

\section{Results}

\subsection{Absorption in the Ice-band}

Figure 4 presents the angularly resolved ice-band SED of T Tau N and S measured on UT 1998 December 11.  By fitting a smooth continuum distribution to the SED, we derive optical depths for the ice feature, $\tau_{ice}$, of 0.1$\pm$0.1 toward T Tau N and 0.5$\pm$0.1 toward T Tau S.  If we consider the total unresolved system, we derive $\tau_{ice}$ = 0.3$\pm$0.1, consistent with Sato et al.'s (1990) value of 0.29$\pm$0.02.  This agreement may, however, be fortuitous considering the photometric variability of the T Tau system and apparent variability of $\tau_{ice}$ (G91; Beck et al. 2000).  Figure 5 shows the ice band SEDs of T Tau S for all three of our observations.  The depth of the ice-band absorption decreased between December 1998 and January 2000 (Table 1).  The observed change in $\tau_{ice}$ is significant at the 2$\sigma$ level and is associated with an increase in the near IR flux (Table 1).  We do not detect a statistically significant change in ice-band optical depth toward T Tau N.
 
It is useful to relate $\tau_{ice}$ and the more familiar extinction in the visible, A$_{v}$.  At this time relationships between $\tau_{ice}$ and A$_{v}$ are available only for stars seen through molecular clouds (Whittet et al. 1988; Teixeira \& Emerson 1999) and their applicability to the denser circumstellar regions considered here is uncertain (see also \S 4).  Using Whittet et al.'s relation,  $\tau_{ice}$ = [m($A_{v}-A_{v}(0)$] where m = 0.093$\pm$0.001 and $A_{v}(0)$ = 3.3$\pm$0.1, we determine A$_{v}$ = 4$\pm$4 mag toward T Tau N and 9$\pm$2 toward T Tau S on 1998 December 11.  On 1999 November 17 and 2000 January 27 the corresponding visual extinctions to T Tau S were 7$\pm$2 and 5$\pm$2 mag.  We cannot use the relation derived by Teixeira \& Emerson because it requires the full width at half maximum of the ice-band absorption and our CVF observations were too widely spaced to measure this.

\subsection{K-band Diagnostics}

The K-band spectrum of T Tau N (Figure 2) shows prominent Br$\gamma$
emission and photospheric absorption lines of Na (2.208
$\mu$m), Ca (2.264 $\mu$m), and the CO $\Delta\nu=$2 bands.
Visible light observations indicate that T Tau N has a
spectral type of K0 and is obscured by A$_{v}\sim$1 (Cohen
\& Kuhi 1979).  J$-$H and
H$-$K colors, derived from photometry of T Tau N
appearing in Kenyon \& Hartmann (1995), also
indicate a value of A$_{v}\sim$1.  
The photospheric absorption features detected in the
K-band are consistent with a K0 spectral type,
although the shallow depth of these features in the
spectrum of T Tau N requires a 2.2$\mu$m continuum excess 
on the order of $>$70\% of the stellar continuum (Figure 2).
The slope of our T Tau N K-band spectrum implies an
A$_V$ of 10$-$12 magnitudes for an early K to
late G type stellar spectrum.  Although complex nebulosity
around T Tau is evident in several studies (e.g. Stapelfeldt
et al. 1998), preferential scattering
of the short wavelength light does not explain the
discrepancy between the different estimates of A$_{v}$,
since the near-IR colors also indicate low extinction.
Rather, it is likely that spectral structure in the infrared excess emission across the K-band is resposible for the observed difference.  These considerations will be discussed in more detail in Prato, Greene \& Simon (in preparation).

Figure 3 shows weak Br$\gamma$ and H$_{2}$ {\it v}=1-0 S(1) emission lines toward T Tau S, at 3$\sigma$ and 2$\sigma$ significance, respectively, but no evidence for H$_{2}$ emission in T Tau N.  Table 2 gives the fluxes of these lines.  Approximately three years after our spectra were obtained Davies et al. (2000) also observed the angularly resolved K-band spectra of T Tau N and S.  Kasper(2000) and Kasper et al. (in preparation) provide a detailed discussion of these spectra.  Their observation was made when T Tau S was in an elevated flux state (G91; Beck et al. 2000), with a 2.2 $\mu$m flux roughly triple the value we measured in October 1996.  They detected Br$\gamma$ emission toward both T Tau N and S.  Their measured  Br$\gamma$ flux of T Tau S, 2.6$\pm$0.3 $\times$ 10$^{-16}$ W m$^{-2}$ s$^{-1}$, is $\sim 2\sigma$ higher than our value (Table 2) suggesting that its Br$\gamma$ flux may vary.

The spectrum of T Tau S presented in Davies et al. (2000) and Kasper (2000) shows no evidence of H$_2$ line emission.  This could indicate that its H$_2$ line emission varied in the three years between our observations.  The H$_2$ emission could, however, arise from an extended circumstellar environment around T Tau S.  In this case, the difference in line fluxes could be attributable to the different solid angles sampled, $\sim (0.8 '')^2$ for our observation, and $(0.25'')^2$ for the Davies et al. spectrum.  High resolution mapping observations are required to resolve this issue.

\section{Discussion}

The projected angular separation of TTau N and S is equivalent to only 100 AU at a distance of 140 pc, but our observations show that there is significantly stronger ice-band absorption towards the IRC.  G91 and Herbst et al. (1997) obtained the same result for absorption in the 10 $\mu$m silicate feature and derived visual extinctions towards T Tau S of 4.6$\pm$2.5 and 5.4$\pm$0.5 mag for the two studies, respectively, consistent with our results in section 3.1.  Our data from the three observations of the water-ice feature suggest an inverse correlation between the near IR flux and ice-band optical depth.  While this has a natural explanation in terms of a decrease of obscuring material along the line of sight, there is a 1/6 probability that this ordering is the result of pairing of random data.  Hence, further observations are required to determine if this correlation is real.

If T Tau S is a young binary viewed through significant obscuration, the typical visual extinctions of 5 to 10 magnitudes derived from ice-band and silicate absorption features are not large enough to account for its observed near IR colors.  K97 estimate that 35 magnitudes of extinction are necessary to redden a 5200 K black-body to $\sim$ 800 K, the characteristic near IR color temperature of T Tau S.  However, 35 magnitudes of extinction should produce enormous ice-band and silicate absorption depths which are not observed.  Two explanations seem possible.  It is likely that the flux that is observed at the line center of the ice and silicate features does not originate from the stellar photosphere, hence the absorption does not trace all of the material that is present toward the system.  Also, whether the size and composition of the dusty material near young stars, and the relations between A$_{v}$ and the ice and silicate optical depths, are the same as in the interstellar medium is unknown and is of considerable interest.

Several possibilities for the geometry of the obscuring material in the T Tau system can account for all or a part of the extinction and its variability observed toward T Tau S.  We may be observing T Tau S through a tenuous extension of the disk of T Tau N, which was detected in HCO$^+$($J = 3-2$) by Hogerheijde et al. (1997), and in 3 mm continuum emission by Akeson et al. (1998).  The extinction variability could be caused by inhomogeneous material in the disk of T Tau N moving through our line of sight, perhaps by Keplerian rotation.  The T Tau N disk would have to lie within the 2:1 resonance to avoid truncation by tidal effects, corresponding to $\sim$1/3 of the semi-major axis if T Tau N and S have equal masses (Artymowicz et al. 1991). This would imply a separation of T Tau N and S much greater than the projected separation of 100 AU.  The obscuring material observed toward T Tau S may also be associated with a nearly edge-on disk around the 0.$''$05 binary, with rotation again responsible for the variability.  Akeson et al. (1998) do not detect any mm continuum emission around T Tau S to a limiting flux of 9 mJy.  However, tenuous material that would escape detection at 3 mm could still be resposible for the extinction toward the T Tau S system.  Another possible explanation for the observed extinction is the episodic embedding and accretion of T Tau S from a disk around the T Tau N - S system (K97).

The observed extinction and variability toward T Tau S could arise from material which is not associated with a disk, but with the complex distribution of gas and dust surrounding the young triple system (Momose et al. 1996; Schuster et al. 1997; Weintraub et al. 1999).  In this case, the absorbing material must have significant structure and large inhomogeneities in absorption on scales less than 100 AU to account for the present small extinction toward T Tau N.  The irrregular dimming of T Tau N by 3-4 visual magnitudes in the early 1900's provides support for this scenario (Lozinskii 1949; Beck \& Simon, in preparation).

There are indications that variations in extinction do not provide the entire explanation for the IR variability of T Tau S and the IRC phenomenon.  The flare discovered by G91 was essentially gray, with a flux increase of $\sim$ 2 magnitudes between 2 and 10 $\mu$m.  Detection of Br$\gamma$ emission associated with T Tau S in both our and the Kasper (2000) spectra indicates continuing accretion activity and evidence for variation in accretion rate.  Additionally, variability of H$_{2}$ line emission would indicate that the flows that shock molecular gas in the environment of T Tau S can vary on a timescale of a few years.

The variability of ice-band optical depth and K-band emission features suggest that the near IR flux variation observed toward T Tau S may be caused by changes in both extinction and accretion rate.  Frequent, simultaneous and high precision measurements of the IR spectral energy distribution, including measurement of the ice and silicate features, and the Br$\gamma$ and H$_{2}$ emission lines, have the potential to clarify the nature of T Tau S and the IRC phenomenon.  Improvements in image quality through the techniques of adaptive optics, and the powerful new IR cameras and spectrometers at the major observatories offer the means for rapid progress.
  
\acknowledgments

We thank the telescope operators at the UH 88$''$ and the IRTF, Chris Stewart, David Griep, Bill Golisch, Charlie Kaminski and Paul Fukumura-Sawada for their assistance with scheduled and service observations.  We are grateful to T. Greene for providing the opportunity to acquire the T Tau K-band spectra.  We thank R. White for helpful discussion, M. Kasper for the link to his Ph.D. thesis and information on his paper in preparation, and the referee for a helpful review.  This research was supported by NSF Grants 94-17191 and 98-19694.

%% The equation environment wil produce a numbered display equation.

%A tree-level amplitude in $e^+e^-$ collisions can be expressed in
%terms of fermion strings of the form
%\begin{equation}
%\bar v(p_2,\sigma_2)P_{-\tau}\hat a_1\hat a_2\cdots
%\hat a_nu(p_1,\sigma_1) ,
%\end{equation}

\clearpage

\begin{table}
\begin{center}
\title{Table 1: Near IR Flux and Ice-Band Properties of T Tau S}
\begin{tabular}{llll}
\tableline\tableline
Date & K & L$'$ & $\tau_{ice}$ \\
 & (Jy)  &  (Jy)  &  \\
%\multicolumn{1}{c}{$P$\tablenotemark{a}} & $P R_{maj}$ & $P R_{min}$ &
%\multicolumn{1}{c}{$\Theta$\tablenotemark{b}} \\
\tableline
1998 Dec 11 & 1.07$\pm$0.08 & 5.5$\pm$0.3 & 0.5$\pm$0.1  \\
1999 Nov 17 & 2.13$\pm$0.05 & 6.0$\pm$0.1 & 0.3$\pm$0.1 \\
2000 Jan 27 & 2.1$\pm$0.2 & 6.9$\pm$0.3 & 0.2$\pm$0.1 \\
\tableline
\end{tabular}
\end{center}
\end{table}

\begin{table}
\begin{center}
\title{Table 2: Br$\gamma$ and H$_{2}$ Line Emission}
\begin{tabular}{llll}
\tableline\tableline
 Species & Wavelength & T Tau N & T Tau S \\ 
 & ($\mu$m) & ($\times$ 10$^{-16}$ W m$^{-2}$ s$^{-1}$) & ($\times$ 10$^{-16}$ W m$^{-2}$ s$^{-1}$)\\
\tableline
H$_{2}$ & 2.1218 & $<$1.5 (3$\sigma$) & 1.1$\pm$0.5 \\
H I & 2.1661 & 7.5$\pm$0.5 & 1.6$\pm$0.5\\
\tableline
\end{tabular}
\end{center}
\end{table}

\clearpage

\appendix

%\section{Appendicial material}




%% Generally speaking, only the figure captions, and not the figures
%% themselves, are included in electronic manuscript submissions.
%% Use \figcaption to format your figure captions. They should begin on a
%% new page.

\clearpage

%% No more than seven \figcaption commands are allowed per page,
%% so if you have more than seven captions, insert a \clearpage

%% There must be a \figcaption command for each legend. Key the text of the
%% legend and the optional \label in curly braces. If you wish, you may
%% include the name of the corresponding figure file in square brackets.
%% The label is for identification purposes only. It will not insert the
%% figures themselves into the document.
%% If you want to include your art in the paper, use \plotone.
%% Refer to the on-line documentation for details.

\figcaption[ttau_fig1.ps]{A 3.05 $\mu$m circular variable filter image of T Tau N and S from observations made on 1998 December 10.  The figure is oriented North up, and East to the left.  The x-y coordinates are in units of 0.$''$028, half pixels.  The contours are 14, 28, 42, 56, 70 and 84\% of the peak flux.  The signal-to-noise is $\sim$ 120 and the flux ratio (South/North) is 0.47$\pm$0.02. Overplotted in the upper right is a cross-cut of the image at an x position of 64.  The apparent peaks at y=$\sim$25, 50, and 75 are the first order Airy rings of the diffraction limited image. \label{Fig. 1}}

\figcaption[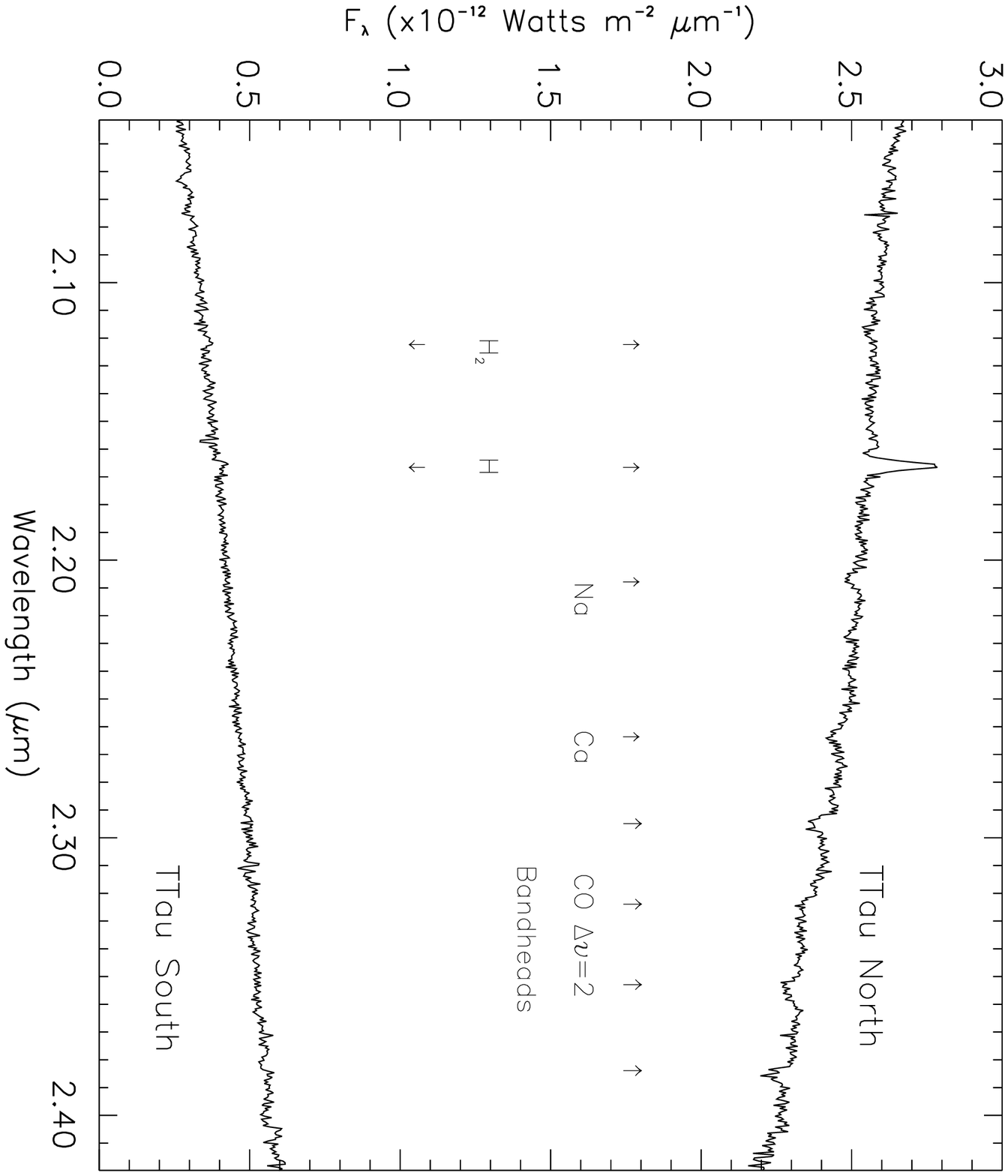]{The angularly resolved K band spectra of T Tau N and S.  The most prominent spectral features are identified. \label{Fig. 2}}

\figcaption[ttau_fig3.ps]{A magnified view of the 2.1 to 2.2 $\mu$m spectral region of T Tau N and S.  H$_{2}$ is detected in emission toward T Tau S, but not toward T Tau N.  Fits to the spectral continua are overplotted.\label{Fig. 3}}

\figcaption[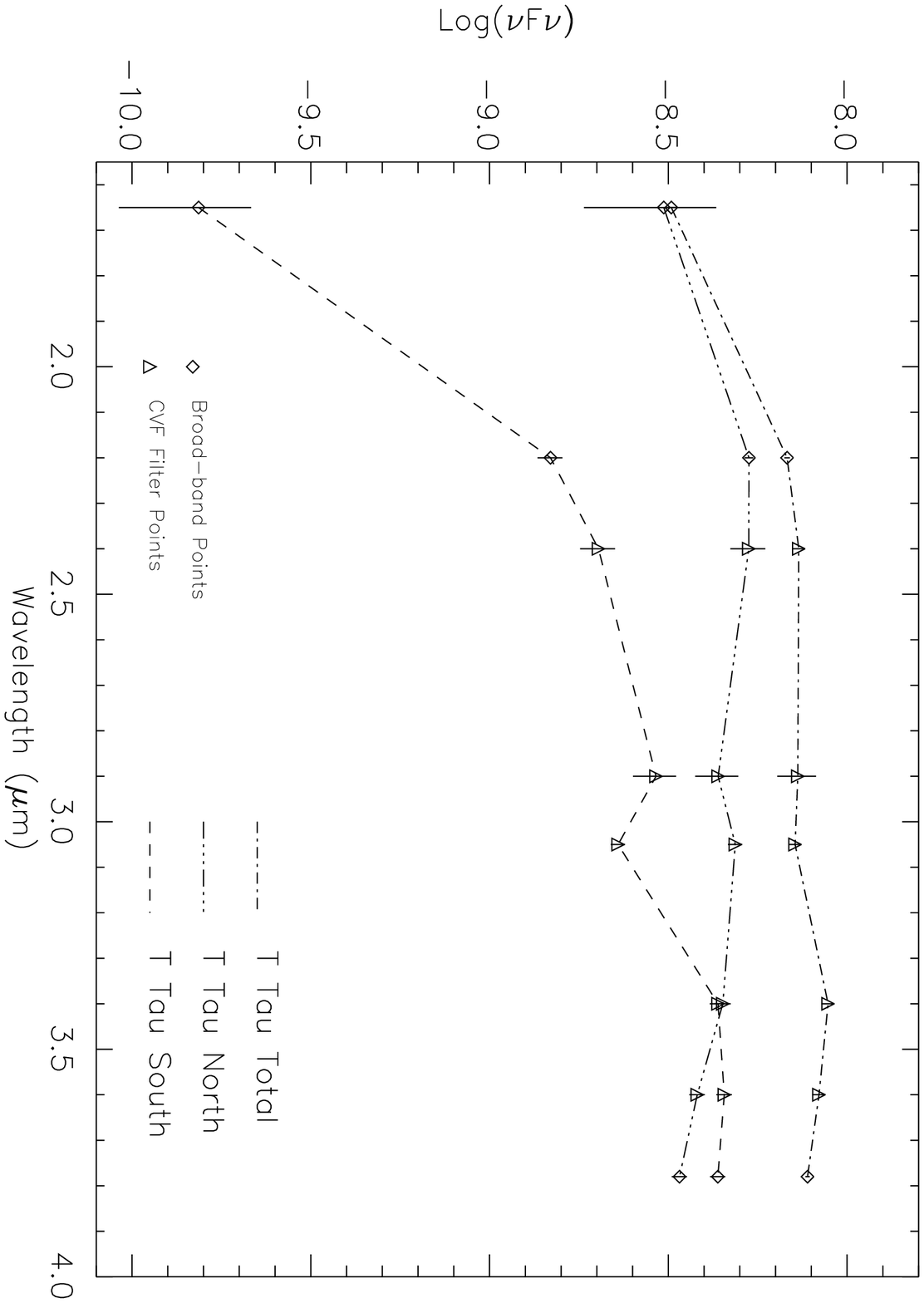]{The total and angularly resolved near IR spectral energy distributions (SEDs) of T Tau N and S from data obtained on 1998 December 11.\label{Fig. 4}}

\figcaption[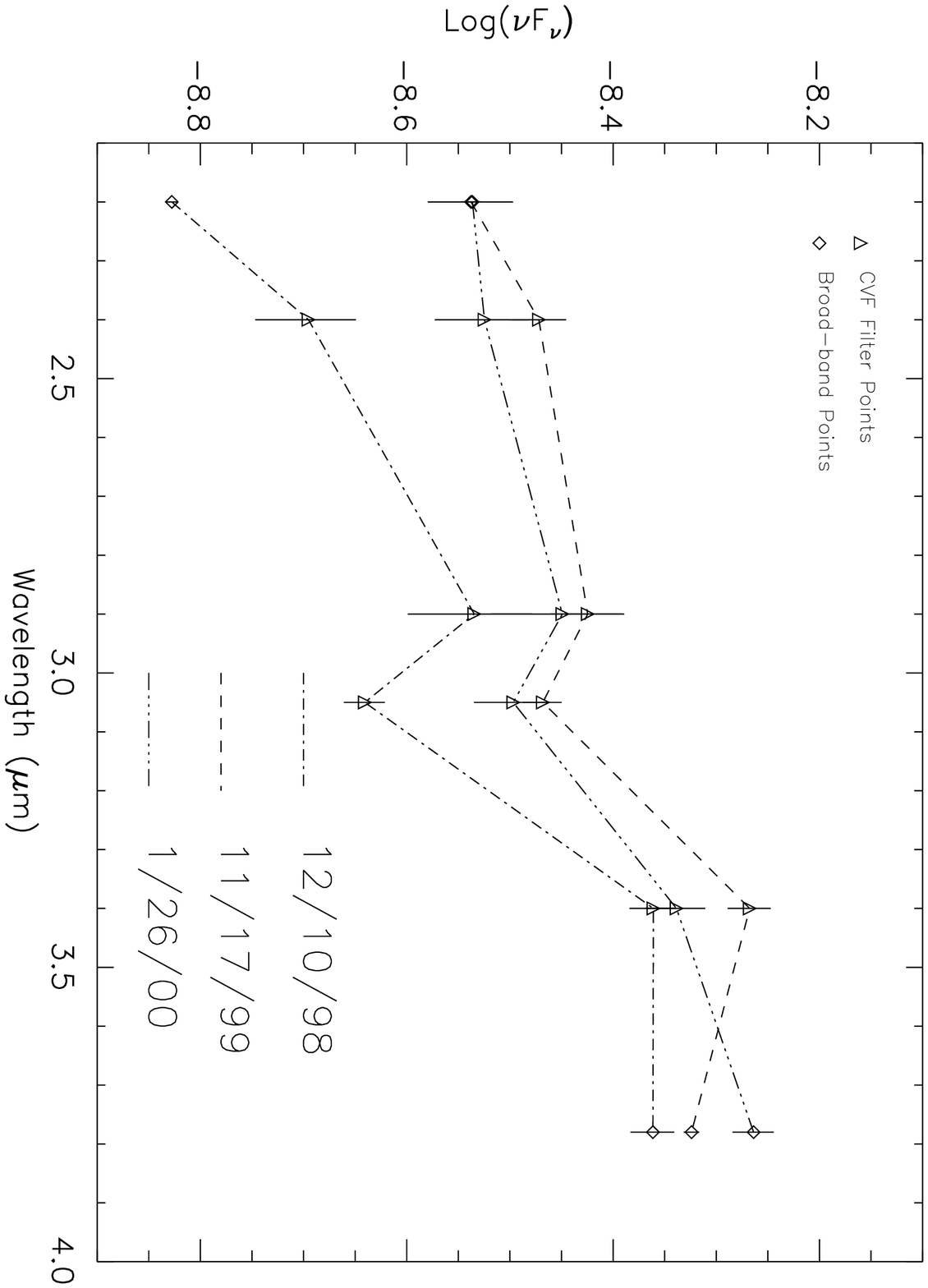]{The observed change in the SED of T Tau S from 1998 December to 2000 January.  The derived ice-band optical depths are listed in Table 1.\label{Fig. 5}}

\begin{figure}
\plotone{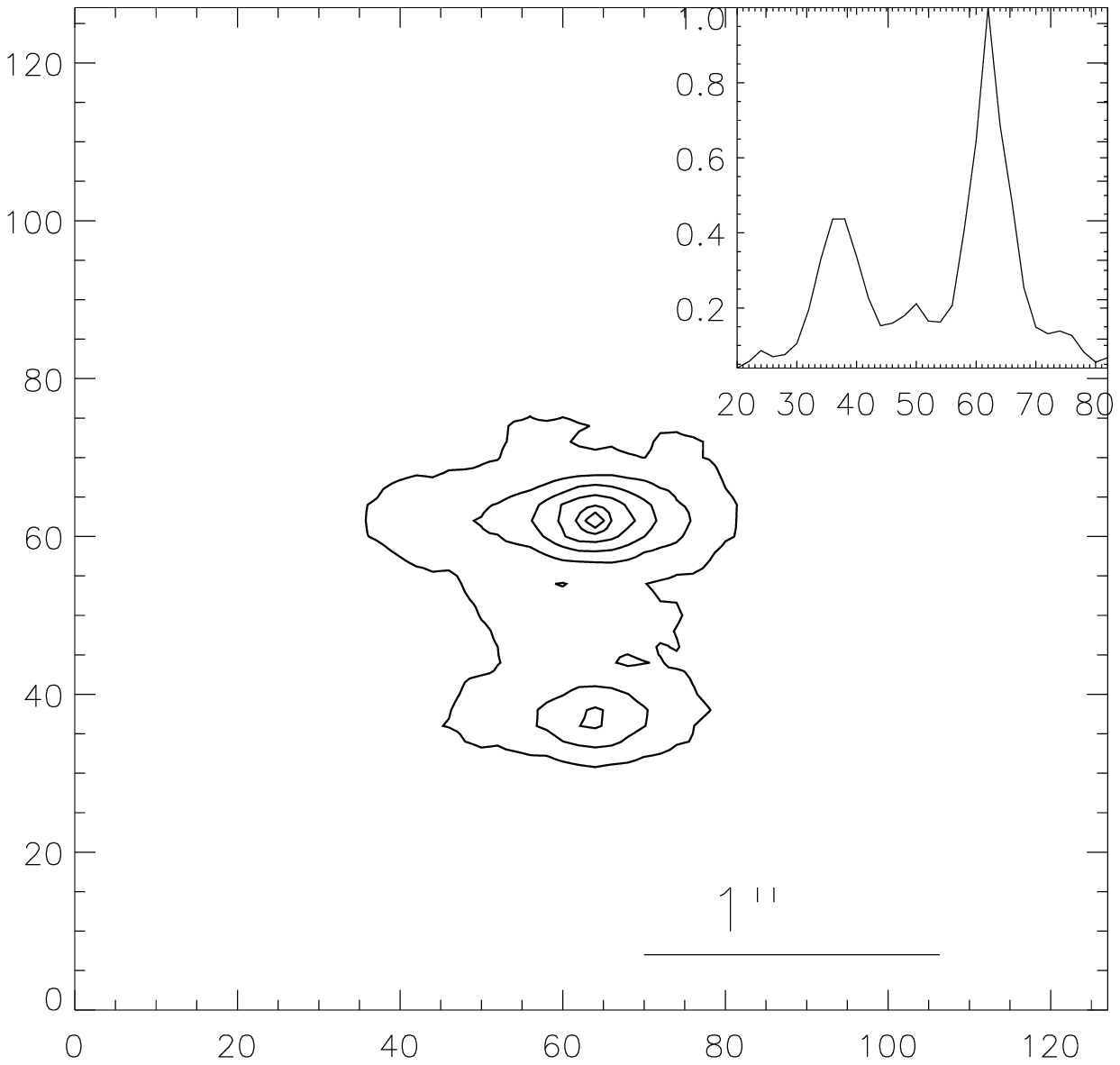}
\end{figure}

\clearpage

\begin{figure}
\plotone{ttau_fig2.ps}
\end{figure}

\clearpage

\begin{figure}
\plotone{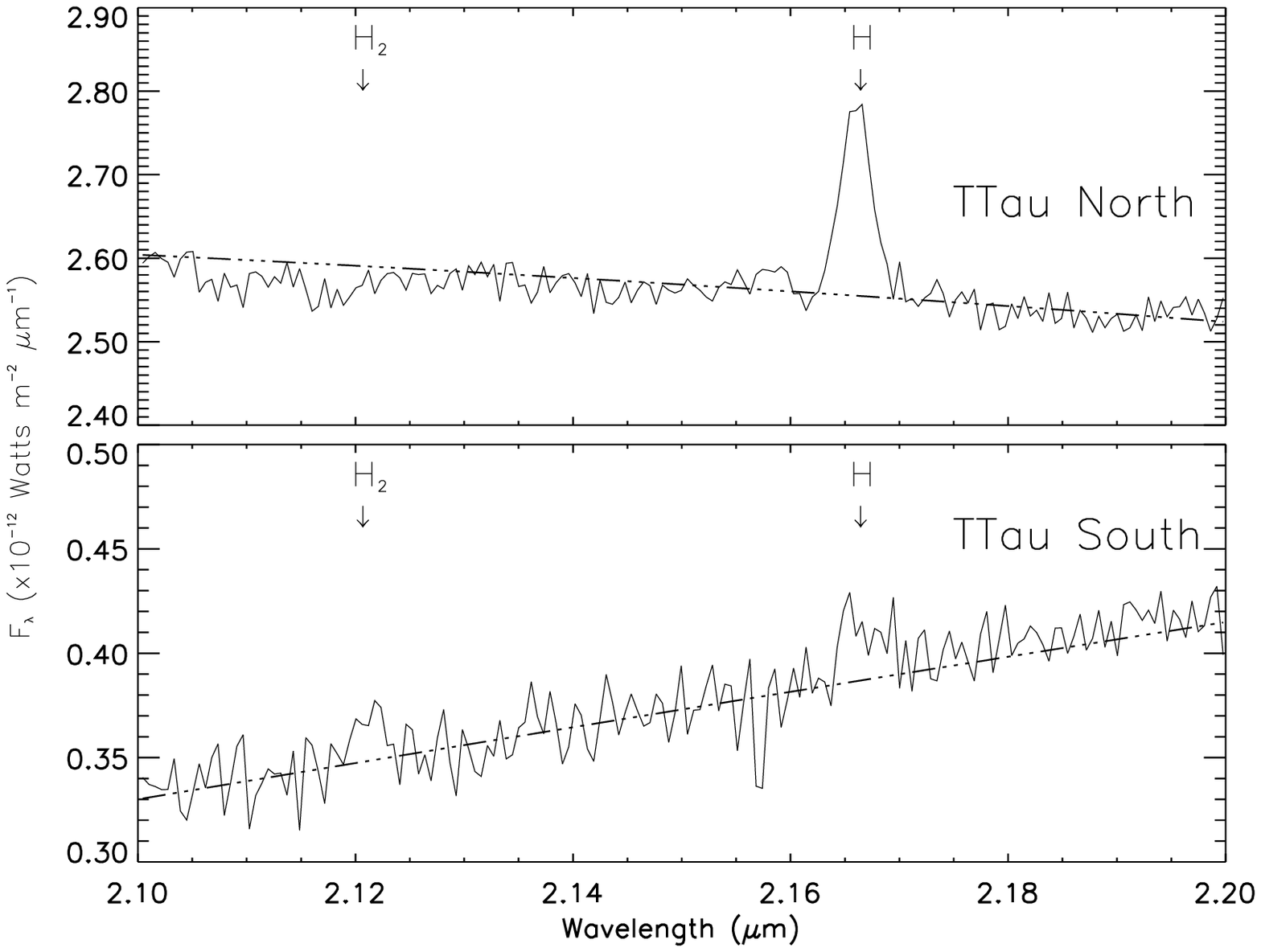}
\end{figure}

\clearpage

\begin{figure}
\plotone{ttau_fig4.ps}
\end{figure}

\clearpage

\begin{figure}
\plotone{ttau_fig5.ps}
\end{figure}

%% The following command ends your manuscript. LaTeX will ignore any text

\end{document}